Phantom chain simulations for the fracture of star polymer networks on the effect of arm molecular weight

Yuichi Masubuchi

Department of Materials Physics, Nagoya University

Nagoya 4648603, JAPAN

mas@mp.pse.nagoya-u.ac.jp

ABSTRACT

This study investigated the fracture of star polymer networks made from prepolymers with various arm molecular weights in the range, $2 \leq N_a \leq 20$, for node functionalities $3 \leq f \leq 8$ and conversion ratios $0.6 \leq \varphi_c \leq 0.95$ by phantom chain simulations. The networks were created via end-linking reactions of star polymers dispersed in a simulation box with a fixed monomer density $\rho = 8$. The resultant networks were alternatively subjected to energy minimization and uniaxial stretch until the break. The stretch at the break, $\lambda_b$, depended on the strand molecular weight $N_s = 2N_a + 1$ with a power-law manner described as $\lambda_b \sim N_s^{0.67}$, consistent with the experiment. However, the strand length before stretch is proportional to $N_s^{0.5}$, which does not explain the observed $N_s$-dependence of $\lambda_b$. The analysis based on the non-affine deformation theory does not interpret the phenomenon either. Instead, the increase of normalized prepolymer concentration concerning the overlapping concentration with increasing $N_s$ explains the result through a rise in the fraction of broken strands.

INTRODUCTION

While the fracture of network polymers has been extensively studied, the fundamental mechanism remains unrevealed[1]. According to earlier studies[2,3], the maximum stretch of a single strand determines the strain at the break of the network. Namely, the maximum stretch $\lambda_{max}$ is written as $\lambda_{max} \sim L/R_0 \sim N^{1/2}$, where $L$ and $R_0$ are the contour length and equilibrium end-to-end distance of the strand, and $N$ is the number of Kuhn segments of the strand. This idea implicitly assumes that the broken strand is included in an aligned path to the elongational direction and that the path comprises only mechanically effective strands. Some simulation studies support these assumptions[4,5].

Meanwhile, some experiments reported interesting results that seem inconsistent with the theory. Akagi et al.[6] determined $\lambda_{max}$ for tetra-PEG gels with various prepolymer molecular weights (which correspond to $N$) and the conversion ratios $\varphi_c$ by applying an extended Gent model[7] to the stress-strain curves. They obtained an empirical relation written as $\lambda_{max} \sim N^{2/3}$. Fujiyabu et al.[8] rationalized this power-law, assuming that prepolymers are in a random, close-packed arrangement, and the strand length before stretch $R_0$ is proportional to $N^{1/3}$, which is different from the unperturbed dimension of the single prepolymer $R_0 \sim N^{1/2}$. Given that $\lambda_{max} \sim R_{max}/R_0$ and $R_{max} \sim N$, $R_0 \sim N^{1/3}$ corresponds to $\lambda_{max} \sim N^{2/3}$. They also demonstrated that the results from both tetra and tri-branch networks exhibit the same scaling concerning the relation between $\lambda_{max}$ and $N$. They argued that the results are unlikely to be due to trapped entanglements owing to their careful preparation of the samples.

It should be noted that the experiments mentioned above did not measure the stretch at the breaking point. Experimentally evaluating network rupture is challenging due to the significant impact of defects at various length scales on the results; thus, few reports can be found on direct measurement of stretch at break for gels and rubbers. Nevertheless, $\lambda_{max}$ obtained from stretch non-linearity may not coincide with stretch at the break because $\lambda_{max}$ corresponds to the average stretch, and the breakage of a small number of strands triggers network rupture.

In this study, the effects of $N$ on the network rupture were systematically investigated for star polymer networks with various node functionalities and conversion ratios through phantom chain network simulations. Since the phantom chain network is employed, neglecting entanglements and osmotic pressure, an intuitive outcome is that the stretch at break $\lambda_b$ is proportional to $N^{1/2}$. However, counterintuitively, the result demonstrated that stretch at break $\lambda_b$ is proportional to $N^{0.67}$, not $N^{1/2}$, consistent with experiment [6]. Further interestingly, the mechanism differs from what was suggested earlier [6]; the power-law $\lambda_b \sim N^{0.67}$ comes from the difference in the prepolymer concentration concerning the overlapping concentration via the increased broken strand fraction. Details are shown below.

MODEL AND SIMULATIONS

Because the model and simulation method are similar to those reported earlier[9,10], only a brief description is given below. Star-branch bead-spring chains replaced prepolymers. There is no interaction between beads except for the connectivity; thus, the chains can cross. These phantom chains were dispersed in a simulation box with periodic boundary conditions and equilibrated by the Brownian dynamics scheme. The equation of motion for the position of bead $\mathbf{R}_i$ is as follows.

$$0 = -\zeta \dot{\mathbf{R}}_i + \frac{3k_B T}{a^2} \sum_k \mathbf{b}_{ik} + \mathbf{F}_i \quad (1)$$

The first term on the right-hand side is the drag force, and $\zeta$ is the friction coefficient. The second term is the balance of tension among connected bonds. $a$ is the average length under equilibrium for the bond vector $\mathbf{b}_{ik} \equiv \mathbf{R}_i - \mathbf{R}_k$, $k_B$ is the Boltzmann constant, and $T$ is the temperature. After sufficient equilibration, gelation via end-linking reactions was simulated[11,12]. According to the experimental setup, the prepolymers were colored in two different chemistries (red and blue), and the reaction occurred only between other colors to eliminate the formation of primary loops.

Similarly to earlier studies[13–16], energy minimization was applied to the obtained network without Brownian motion for the total bond energy written below.

$$U = \frac{3k_BT}{2a^2} \sum_{i,k} \mathbf{b}_{ik}^2 \qquad (2)$$

The energy-minimized network was elongated quasi-statically by alternative application of affine elongation with an infinitesimal strain $\Delta\varepsilon$ and energy-minimization. During the elongation, bonds are removed when they exceed a specific maximum value, $b_c$. The simulation continued until the network percolation was eliminated to the elongated direction. Elongational (true) stress $\sigma$ was recorded as a function of $\varepsilon$, and strain and stress at break, $\varepsilon_b$ and $\sigma_b$, were obtained from such a plot. Work for fracture $W_b$ was calculated by numerically integrating the stress-strain relation until the network breaks. The employed quasi-static scheme avoids the effects of stretch rate, neglecting structural relaxations after every bond scission[17].

The simulations were performed with nondimensionalized quantities, and units of length, energy, and time were chosen at $a$, $k_BT$, and $\tau = \zeta a^2/k_BT$. For star prepolymers, $f$ and $N_a$ were varied as $3 \leq f \leq 8$ and $2 \leq N_a \leq 20$. The number of beads in the system was ca. 35,000, slightly depending on $f$ and $N_a$ since the number of prepolymers must be integer. The bead number density $\rho$ was fixed at 8, and thus, the polymer concentration normalized by the overlapping concentration[18] $c/c^*$ ranges depending on $f$ and $N_a$, as shown in SI (Fig S1). For the Brownian dynamics, a second-order integration scheme[19] was employed, with the step size of $\Delta t = 0.01$. The parameters for the end-linking reaction were chosen according to the previous studies[11,12]; the critical distance was $r_r = 1$, and the reaction probability was $p_r = 0.1$. During the gelation, snapshots at $\varphi_c = $ 0.6-0.95 were stored for stretch. The energy minimization scheme was the Broyden-Fletcher-Goldfarb-Sanno method[20]. The bead displacement parameter was $\Delta r = 0.01$, and the energy conversion parameter was $\Delta u = 10^{-4}$. The magnitude of elongation at the stepwise stretch was $\Delta \varepsilon = 0.01$. $b_c$ was fixed at $\sqrt{1.5}$ unless stated. For statistics, eight independent simulation runs were performed for each set of parameters $(f, N_a, \varphi_c)$. Typical snapshots are shown in SI (Figs S2 and S3), where no structural inhomogeneity was observed even under elongation. One may argue that the network fragmentation seen in the snapshots implies flaws in the simulation scheme. However, the results below are essentially unchanged for a different scheme disallowing the fragmentation, as shown later.

RESULTS AND DISCUSSION

The stress-strain relations were recorded during the elongation process, as shown in SI (Fig S4), and strain at the break $\varepsilon_b$, stress at the break $\sigma_b$, and the work for fracture $W_b$ were obtained. $\varepsilon_b$ is converted to stretch at break $\lambda_b \equiv \exp(\varepsilon_b)$ and shown in Fig 1 for $\varphi_c = 0.95$ against the strand molecular weight $N_s = 2N_a + 1$. (Note that a single strand consists of a pair of star arms with $N_a$ bonds and one linking bond.) Figure 1 (a) exhibits a power-law dependence of $\lambda_b$ on the strand molecular weight described as $\lambda_b \propto N_s^\alpha$. For this exponent $\alpha$, $\lambda_b$ is normalized by $N_s^{1/2}$ in Fig. 1 (b). The result exhibits that $\alpha \sim 0.67 > 0.5$. This exponent is close to Akagi et al.[6], who suggested that the average strand length before stretch $R_0$ may deviate from the expected relation $R_0 \propto N_s^{1/2}$. However, the mechanism of this phenomenon in this case is different, as discussed later.

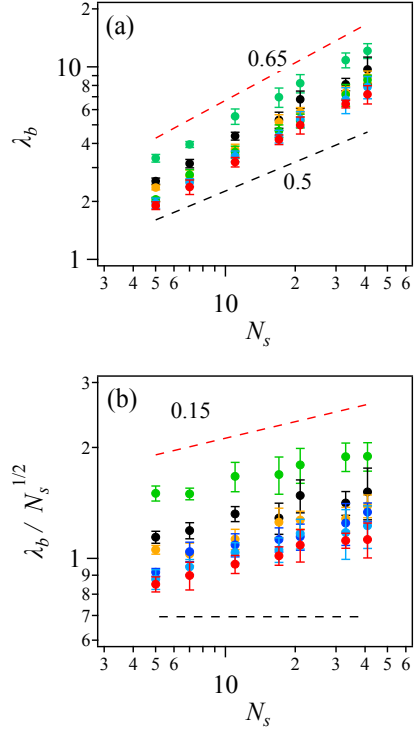

**Figure 1** Stretch at break $\lambda_b \equiv \exp(\varepsilon_b)$ plotted against strand molecular weight $N_s$ at $\varphi_c = 0.95$ (a) and normalized by $N_s^{1/2}$ (b) for $f = 3$ (green), 4 (black), 5 (orange), 6 (blue), 7 (cyan), and 8 (red), respectively. Error bars indicate standard deviations for eight simulation runs. Black and red broken lines indicate power-law relations with the exponent of 0.5 and 0.67, respectively.

Note that this study employed the linear spring differently from earlier studies[9,10,21–26], in which the spring constant $f_{ik} = (1 - \mathbf{b}_{ik}^2/b_{\max}^2)^{-1}$ was used. The reason is to eliminate possible effects of nonlinearity in the mechanical response and to reduce the number of simulation parameters. Nevertheless, the conclusion for the impact of $N_s$ was not affected, at least for the case of $b_{\max} = 2$, as shown in SI (Fig S5).

In Figs. 2 (a) and (b), $\sigma_b$ and $W_b$ are shown as functions of $N_s$. These values increase with increasing $N_s$ and hit saturation for large-$N_s$. The effects of $f$ are not apparent. These results are partly due to the branch point density $v_{br} \equiv \rho/(fN_a + 1)$, which decreases with increasing $f$ and $N_a$ since the bead number density $\rho$ is fixed at 8. To eliminate the effect of $v_{br}$, Figs. 2 (c) and (d) display $\sigma_b/v_{br}$ and $W_b/v_{br}$, which systematically increase as $f$ increases. $N_s$-dependence also becomes intense. For $N_s \geq 20$, power-law-like relations are seen with a slope of 1.2, as shown by broken red lines. However, the entire behavior is not described by a power-law.

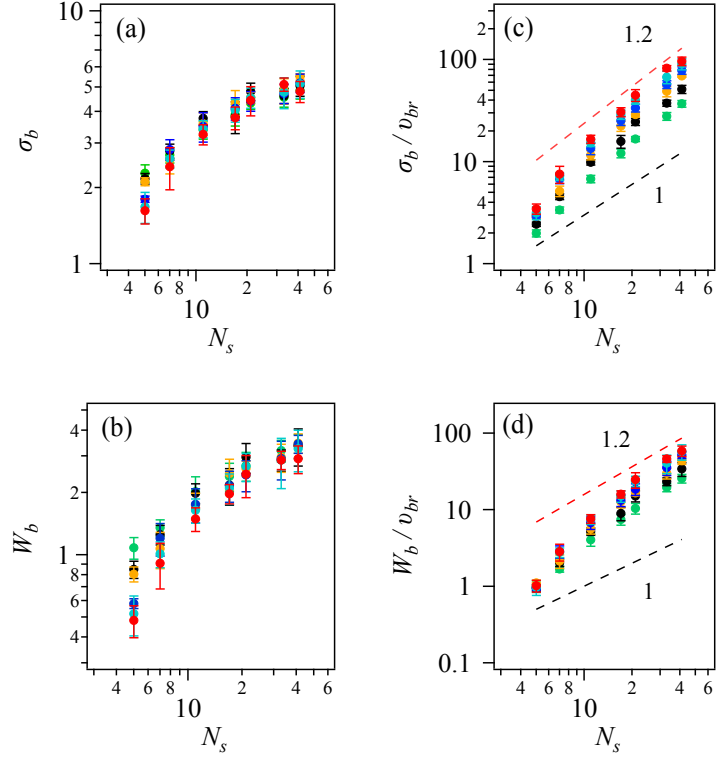

**Figure 2** Stress at break $\sigma_b$ (a) and work for fracture $W_b$ (b) plotted against strand molecular weight $N_s$ at $\varphi_c$ =0.95 with various $f$ values and those normalized by the branch point density $\upsilon_{br}$ shown in (c) and (d). $f = 3$ (green), 4 (black), 5 (orange), 6 (blue), 7(cyan), and 8 (red), respectively. Error bars indicate standard deviations for eight simulation runs. Broken black and red lines indicate power-law relations with the exponents of 1 and 1.2, respectively, for eye-guide.

One may argue that the abovementioned $N_s$−dependence of fracture characteristics might be specific for the case with $\varphi_c$ =0.95. Concerning this issue, the fracture characteristics obtained for different $\varphi_c$ and $f$ values are summarized as functions of $\xi$, the number of the minimum closed loops per branch point in the percolated network. (See earlier studies[1,27,28] for details of $\xi$.) Previous studies[10,23–26] reported that $\lambda_b$, $\sigma_b/\upsilon_{br}$, and $W_b/\upsilon_{br}$ for different $\varphi_c$ and $f$ values draw master curves if plotted against $\xi$. As shown in SI (Fig S6), $\xi$ for the examined networks is consistent with the mean-field theory[29,30], demonstrating that the network connectivity is random. Nevertheless, the effects of $N_s$ shall be discussed for these curves below.

In Figure 3 (a), $\lambda_b$ for various $N_a$ are plotted against $\xi$. For small $\xi$, error bars are relatively large since the data were taken for the networks with small $\varphi_c$ and $f$ values, and distribution among different simulation runs is larger than the cases with large $\varphi_c$ and $f$ values. Nevertheless, $\lambda_b$ obtained for various $\varphi_c$ and $f$ values lie on a master curve for each $N_a$, and the master curve systematically shifts upwards as $N_a$ (thus $N_s$) increases, without changing the slope. This behavior can be approximately described as $\lambda_b = A_\lambda \xi^{-0.3}$ and the obtained $A_\lambda$ values via the fitting are shown in Fig 3 (b) as functions of $N_s$, demonstrating that $A_\lambda \propto N_s^{0.67}$, in harmony with the result in Fig 1.

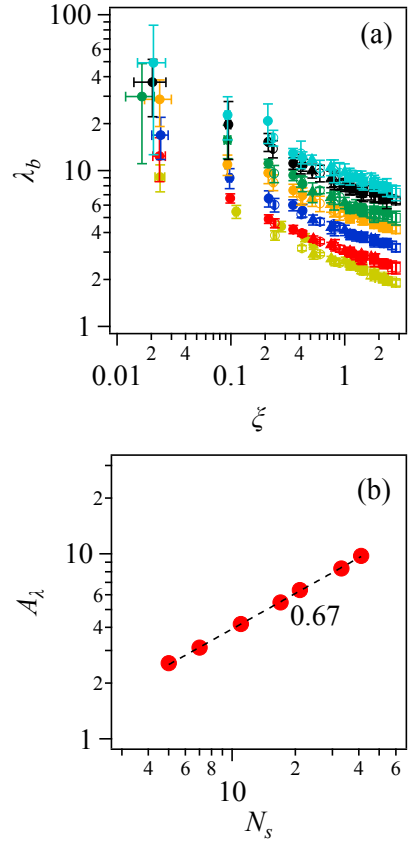

**Figure 3** Stretch at break $\lambda_b$ plotted against $\xi$ for various $N_a$, $f$, and $\varphi_c$ (a). $N_a$ values are 2 (yellow), 3 (red), 5 (blue), 8 (orange), 10 (green), 16 (black), and 20 (cyan). Symbols indicate $f = 3$ (filled circle), 4 (unfilled circle), 5 (filled triangle), 6 (unfilled triangle), 7 (filled square), and 8 (unfilled square). Error bars correspond to the standard deviations for eight different simulation runs for each condition. The $\xi$-dependence of $\lambda_b$ is empirically described by a power-law function $\lambda_b = A_\lambda \xi^{-0.3}$, and the parameter $A_\lambda$ is shown as a function of $N_s$ in Fig (b). The broken line indicates a slope of 0.67.

In Figures 4(a) and (b), $\sigma_b/\nu_{br}$ and $W_b/\nu_{br}$ are shown as functions of $\xi$. Like $\lambda_b$, these quantities obtained from various $f$ and $\varphi_c$ draw a master curve for each $N_a$, and the curves exhibit upward shifts as $N_a$ increases. The results in (a) and (b) are empirically fitted by power-law functions $\sigma_b/\nu_{br} = A_\sigma \xi^{\alpha_\sigma}$ and $W_b/\nu_{br} = A_W \xi^{\alpha_W}$, and the fitting parameters are shown in (c) and (d) against $N_s$. The $N_s$-dependences of $A_\sigma$ and $A_W$ are consistent with Fig 2, in which $\sigma_b/\nu_{br}$ and $W_b/\nu_{br}$ increase with increasing $N_s$ but not in power-law. Concerning the exponents $\alpha_\sigma$ and $\alpha_W$, they are roughly 0.5 and 0.3, although $\alpha_W$ for $N_a = 2$ ($N_s = 5$) is apparently smaller than the others.

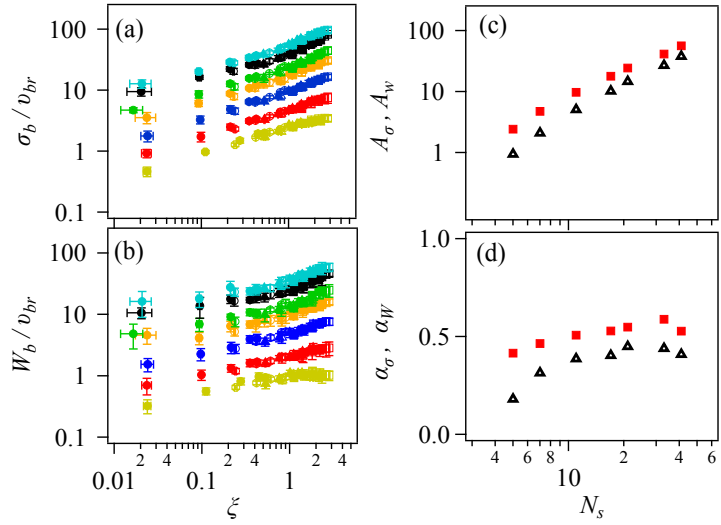

**Figure 4** Stress at break $\sigma_b$ (a) and work for fracture $W_b$ (b) normalized by the branch point density $\nu_{br}$ plotted against $\xi$ for various $N_a$, $f$, and $\varphi_c$. $N_a$ values are 2 (yellow), 3 (red), 5 (blue), 8 (orange), 10 (green), 16 (black), 20 (cyan), and 32 (violet). Symbols indicate $f = 3$ (filled circle), 4 (unfilled circle), 5 (filled triangle), 6 (unfilled triangle), 7 (filled square), and 8

(unfilled square). Error bars correspond to the standard deviations for eight different simulation runs for each condition. The $\xi$-dependences of $\sigma_b/v_{br}$ and $W_b/v_{br}$ are empirically described by power-law functions $\sigma_b/v_{br} = A_\sigma \xi^{\alpha_\sigma}$ and $W_b/v_{br} = A_W \xi^{\alpha_W}$ and the parameters are shown as functions of $N_s$ in Figs (c) and (d).

From this point forward, let us concentrate on the dependence of $\lambda_b$ on $N_s$. In the earlier studies[6,8], the power-law exponent was attributable to the strand length before stretch $R_0$, as mentioned in the introduction, and the earlier conjecture was $R_0 \sim N_s^{1/3}$. Figure 5 (a) displays the ensemble average $\langle R_0 \rangle$ plotted against $N_s$ for various $f$ at $\varphi_c = 0.95$. The power-law exponent is close to 1/2, rather than 1/3, compared to the slopes for eye-guide. Figure 5 (b) shows the distribution of $R_0$ at $(f, \varphi_c) = (4, 0.95)$ for various $N_s$ normalized by $N_s^{-1/2}$. This plot also demonstrates that $R_0 \propto N_s^{1/2}$ for the entire range of $R_0$. One may argue that $R_0$ examined here is after energy minimization and different from what was discussed earlier. Indeed, as reported previously, energy minimization reduces the strand length, as shown by broken curves in Fig 5 (b). However, the scaling of $R_0$ concerning $N_s$ is unaffected. Nevertheless, the results in Fig 5 imply that $R_0$ does not follow $R_0 \sim N_s^{1/3}$, and does not explain the relation $\lambda_b \propto N_s^{0.67}$.

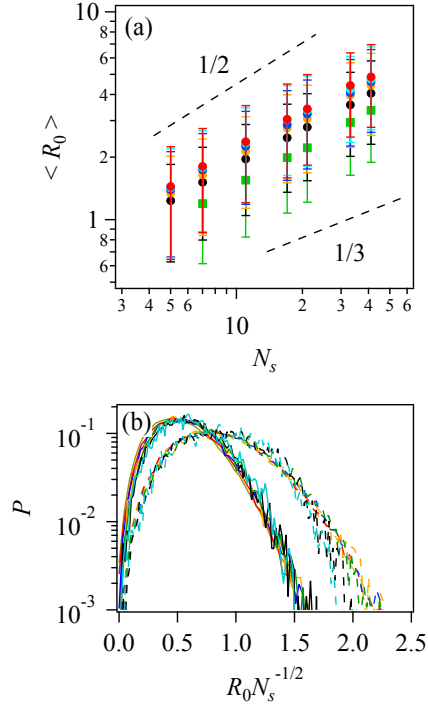

**Figure 5** Average strand length before stretch after energy minimization $\langle R_0 \rangle$ plotted against $N_s$ (a) for various $f$ at $\varphi_c = 0.95$ (a), and the distribution of $R_0$ at $(f, \varphi_c) = (4, 0.95)$ for various $N_s$ normalized by $N_s^{-1/2}$. In Fig 4 (a), $f = 3$ (green), 4 (black), 5 (orange), 6 (blue), 7 (cyan), and 8 (red), respectively. Error bars show the standard deviation of $R_0$. Broken lines indicate slopes of 1/2 and 1/3 for eye-guide. In Fig 4 (b), $N_a$ values are 2 (yellow), 3 (red), 5 (blue), 8 (orange), 10 (green), 16 (black), and 20 (cyan). Solid and broken curves are after and before energy minimization.

Another possibility is the non-affine deformation of the strands. According to the phantom network theory by Panyukov and Rubinstein [31,32], strand deformation in phantom chain networks is suppressed compared to the affine deformation as follows. (See eq 6 in Ref 31.)

$$\langle R^2 \rangle = \langle R_0^2 \rangle \frac{n + \lambda^2 N_s}{n + N_s} \qquad (3)$$

Here, $n$ is the parameter describing the magnitude of deviation from the background affine deformation. Figure 6 (a) shows $\langle R^2 \rangle$ against $\lambda$ for $(f, \varphi_c) = (4, 0.95)$ with various $N_a$. The simulation results shown by symbols follow eq 3 in the region $\lambda \leq 2$ shown by broken black curves. As stretched, the data exhibit upward deviation from these curves. However, before bond breakage starts, if different parameter values are employed, the deviated data can also be fitted by eq 3. See the broken red curves. From such plots, the parameter $n$ was determined by the fitting and shown in Fig 6 (b) as a function of $N_s$. As seen in Fig 6 (a), $n$ changes depending on $\lambda$, and the value for small $\lambda$ is larger than that for large $\lambda$. Nevertheless, the plot demonstrates linear relations $n = cN_s$ for both cases irrespective of $f$. Since $\langle R_0^2 \rangle \propto N_s$ as shown in Fig. 5, under small deformations eq. 3 is rewritten as $\langle R^2 \rangle = \langle R_0^2 \rangle N_s (c + \lambda)/(c + 1) \propto N_s$. For large deformations, the parameter $\langle R_0^2 \rangle$ must be modified to fit the data, and shown in Fig 6 (c) as $\langle R'_0{}^2 \rangle$. This plot demonstrates that $\langle R'_0{}^2 \rangle$ is roughly described by $N_s^\gamma$ with the exponent $\gamma \geq 1$, slightly depending on $f$. One may argue that this nonlinearity of $\langle R'_0{}^2 \rangle$ against $N_s$ can explain the $N_s$-dependence of $\lambda_b$ in Fig 1. However, $\langle R'_0{}^2 \rangle \propto N_s^\gamma$ with $\gamma \geq 1$ contradicts $\lambda_b \propto N_s^{0.67}$ according to the single strand idea, in which $\lambda_b \propto R_{max}/R'_0$ and $R_{max} \propto N_s$ are assumed; these assumptions give $\lambda_b \propto N_s/N_s^{\gamma/2} = N_s^{(1-\gamma/2)}$, and the exponent is smaller than 0.5 when $\gamma \geq 1$, which is inconsistent with Fig 1. These results imply that the non-affine deformation does not explain the $N_s$-dependence of $\lambda_b$.

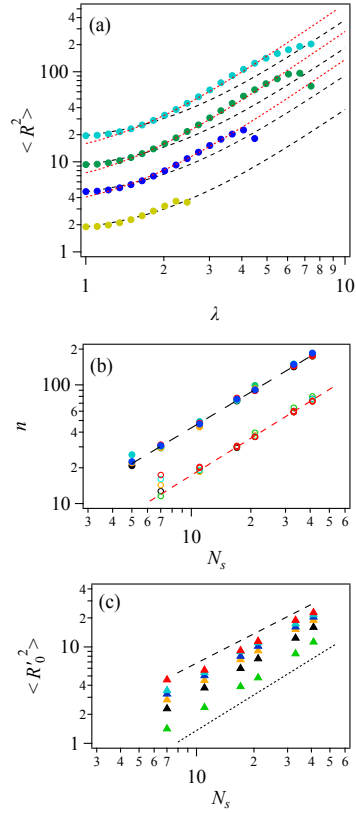

**Figure 6** Average strand length $\langle R^2 \rangle$ under elongation as a function of stretch $\lambda$ (a) and fitting parameters to eq 3 (b and c) for the networks with $\varphi_c = 0.95$. In Fig (a), $f = 4$ and $N_a$ values are 2 (yellow), 5 (blue), 10 (green), and 20 (cyan). The broken black and red curves show eq 3 fitted for small and large $\lambda$ ranges. In Fig (b), the parameter $n$ in eq 3 obtained from the data in Fig (a) via fitting plotted against $N_s$ for small (filled symbols) and large (unfilled symbols) deformations. The broken black and red lines indicate the relations $n = 4.3N_s$ and $n = 1.9N_s$, respectively. In Fig (c), the parameter $\langle R_0^2 \rangle$ under the large $\lambda$ range obtained from the data in Fig (a) via fitting to eq 3 is shown as $\langle R'_0{}^2 \rangle$ against $N_s$. The black broken and dotted curves indicate slopes of 1 and 1.2, respectively. In Figs (b) and (c), $f$ values are 3 (green), 4 (black), 5 (orange), 6 (blue), 7 (cyan), and 8 (red), respectively.

It so appears that the relation $\lambda_b \propto N_s^{0.67}$ cannot be explained by the single strand extension. Wang et al.[33] mentioned that unbroken strand contribute to breakage of connected strands. Concerning this issue, Fig 7 shows the ratio of broken strands $\varphi_{bb}$. As shown in Fig 7 (a), for $\varphi_c = 0.95$ with various $f$, $\varphi_{bb}$ increases with increasing $N_s$, explaining the network toughness depending on $N_s$. The increase of $\varphi_{bb}$ with increasing $N_s$ is seen irrespective of $\varphi_c$, as demonstrated in Figure 7 (b), where $\varphi_{bs}/v_{br}$ is plotted against $\xi$. As reported previously, $\varphi_{bs}/v_{br}$ obtained for various $f$ and $\varphi_c$ values lie on a master curve depending on $N_a$. The $N_a$ dependence of the master curve is consistent with Fig 7 (a), showing upward shifts as $N_a$ increases. This $N_a$ dependence of $\varphi_{bs}$ reflects the maturity of networks. Even though $\xi$ is consistent with the mean-field theory irrespective of $N_a$, some cycles do not effectively contribute stress. Figure 7 (c) shows the modulus $G$ divided by $v_{br}$ plotted against $\xi$. If all the cycles play their roles equivalently, $G/\nu_{br} = \xi$, according to the phantom network theory. (Note that $\xi$ in this study is cycle rank per branch point, and the cycle rank density is written as $\xi\nu_{br}$.) This theoretical description is shown by the dotted line in Fig 7(c), and $G/\nu_{br}$ shows downward deviation from that line, as $N_a$ decreases. As reported previously[26], $G/\nu_{br}$ corresponds to the effective cycle rank (per branch point) $\xi_{eff}$, and that depends on the prepolymer density. In this study, where the segment density $\rho$ is fixed for all the cases, $c/c^*$ depends on $N_a$, as shown in SI (Fig S1). For instance, for $N_a = 2$, $c/c^*$ ranges between 1.9 and 4.2, whereas from 5.8 to 12.2 for $N_a = 20$. Although no theoretical expression can be given, the effect of $c/c^*$ on network toughness can explain $N_s$-dependence of $\lambda_b$ observed with fixed $\rho$, at least partly,

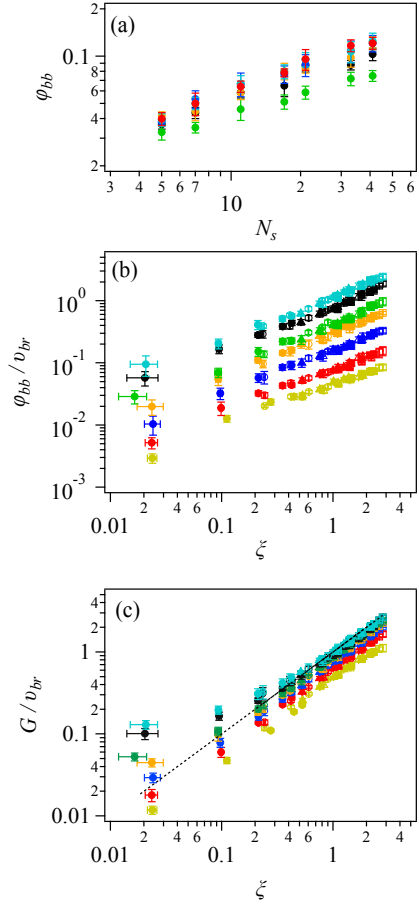

**Figure 7** The fraction of broken bonds $\varphi_{bs}$ for $\varphi_c = 0.95$ with $f$=3 (green), 4 (black), 5 (orange), 6 (blue), 7(cyan), and 8 (red), plotted against $N_s$ (a), and $\varphi_{bs}/v_{br}$ (b) and $G/v_{br}$ as functions of $\xi$ for $N_a$= 2 (yellow), 5 (blue), 10 (green), and 20 (cyan). Error bars correspond to the standard deviations for eight different simulation runs for each condition. The dotted line in Fig (c) shows the relation $G/v_{br} = \xi$. For Figs (b) and (c), symbols indicate $f = 3$ (filled circle), 4 (unfilled circle), 5 (filled triangle), 6 (unfilled triangle), 7 (filled square), and 8 (unfilled square).

For completeness, possible artifacts in the model and simulation scheme employed are discussed in SI (Figs S7- S9). The data demonstrate that the simulation parameters $\Delta u$, $\Delta r$, and $b_c$ do not affect the abovementioned main results. In particular, the simulations starting from the same network structure were performed by changing the value of $b_c$, and the expected relations $\lambda_b \propto b_c$ and $\sigma_b \propto b_c^2$ were confirmed in Fig. S8. Concerning the network fragmentation, a different simulation scheme was examined to break only one bond with the maximum bond length at a time to disallow the fragmentation. See Fig. S9 (a)-(d) for the snapshots. The stress-strain relation (Fig. S9 (e)) was affected by this change only slightly, and thus, the relation $\lambda_b \propto N_s^{0.67}$ was maintained, as shown in Fig. S9 (f) for $\varphi_c = 0.95$. The changes in the fracture characteristics were within errors, as shown in Fig. S10. The simulation with Brownian motion exhibits $\lambda_b \propto N_s^\alpha$ with $\alpha > 1/2$ in Fig S11.

CONCLUSIONS

By phantom chain network simulations, the effects of prepolymer arm molecular weight $N_a$ for star polymer networks on the fracture under stretch were investigated with various node functionalities $f$ and conversion ratios $\varphi_c$. The results obtained were analyzed regarding the effects of the network strand molecular weight $N_s = 2N_a + 1$. For the networks for which $f$ and $\varphi_c$ are common, stretch at break $\lambda_b$ exhibits the relation $\lambda_b \propto N_s^{\alpha}$, and the exponent $\alpha$ is larger than 1/2, consistent with the experimental report. The origin of this discrepancy was analyzed in terms of the strand length before stretch $R_0$. The results indicate that $R_0$ is scaled by $N_s^{1/2}$ including the distribution, differently from the earlier explanation in which $R_0$ scaling is assumed as $N_s^{1/3}$. The strand deformation is also analyzed according to the non-affine deformation theory, but the single-strand deformation does not explain the $N_s$-dependence of $\lambda_b$. The $N_s$-dependences of the fraction of broken strands $\varphi_{bs}$ and the modulus $G/\nu_{br}$ imply that the effect of $N_s$ comes from the effective cycle rank that depends on $c/c^*$.

Although the significance of cycle rank is obvious, other structural parameters like minimum path length[4,5], loop density[34,35], loop length[33,36], etc., may be better to construct theoretical models. It should also be noted that the reported results are for energy-minimized phantom chain networks. Thus, the results may vary if one considers osmotic force, excluded volume interactions and entanglements, etc. Subsequent studies in such directions are ongoing, and the results will be reported elsewhere.

SUPPORTING INFORMATION

The supporting information includes (i) the numerical data table for the fracture characteristics for all the examined systems, (ii) $c/c^*$ for the examined prepolymer systems, (iii) simulation snapshots and stress-strain curves, (iv) $\xi$ of examined networks compared to the mean-field theory, and (v) assessment of the employed simulation scheme.

ACKNOWLEDGEMENTS

This study is partly supported by JST-CREST (JPMJCR1992) and JSPS KAKENHI (22H01189).

TOC Graphics

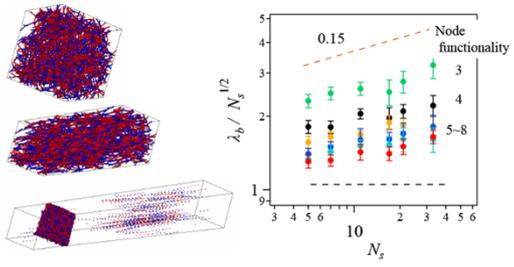

Supporting information for "Phantom chain simulations for the fracture of star polymer networks on the effect of arm molecular weight"

Yuichi Masubuchi
Department of Materials Physics, Nagoya University
Nagoya 4648603, JAPAN
mas@mp.pse.nagoya-u.ac.jp

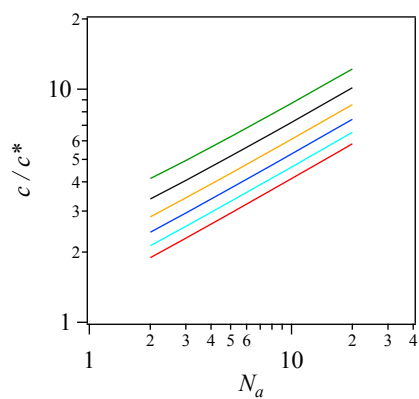

**Figure S1** Polymer concentration in sols normalized by the overlapping concentration for $f = 3$ (green), 4 (black), 5 (orange), 6 (blue), 7 (cyan), and 8 (red) plotted against $N_a$ at the bead concentration $\rho = 8$.

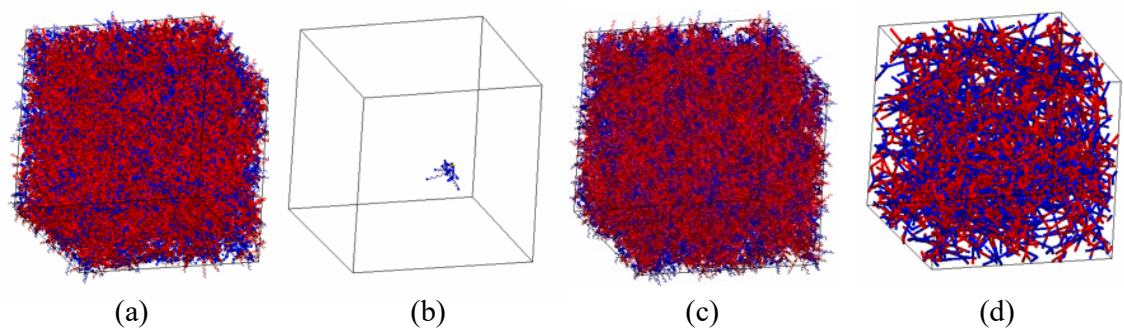

**Figure S2** Typical snapshots of the simulation for a system containing 2240 prepolymers with $N_a = 5$ and $f = 4$ at $\rho = 8$ (a), a single prepolymer in the sol (b), after gelation of the melt at $\varphi_c$=0.95 (c), and after energy minimization before stretch (d). Two types of chemistries are shown in blue and red, respectively.

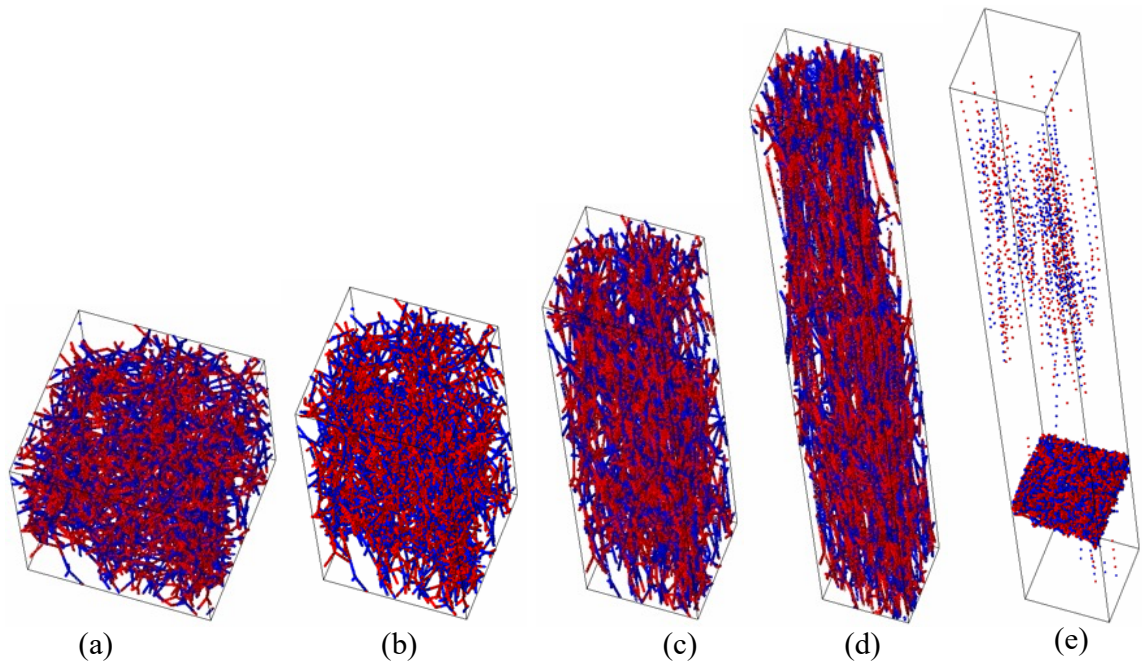

**Figure S3** Typical snapshots under stretch for the network shown in Fig S1 ($N_a = 5$, $f = 4$, and $\varphi_c = 0.95$) at the Henchy strain $\varepsilon$ of 0 (a), 0.5 (b), 1.0 (c), 1.5 (d), and 1.54 (e). Dots seen in Fig (e) are fragments separated from the main domain.

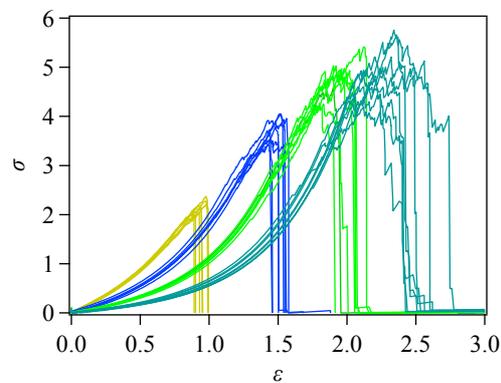

**Figure S4** Typical stress-strain curves for $(f, \varphi_c) = (4, 0.95)$. Yellow, blue, green, and cyan curves are the results for $N_a = 2, 5, 10,$ and $20$, respectively. Eight independent simulation runs were conducted for each condition, and each curve displays an individual simulation run.

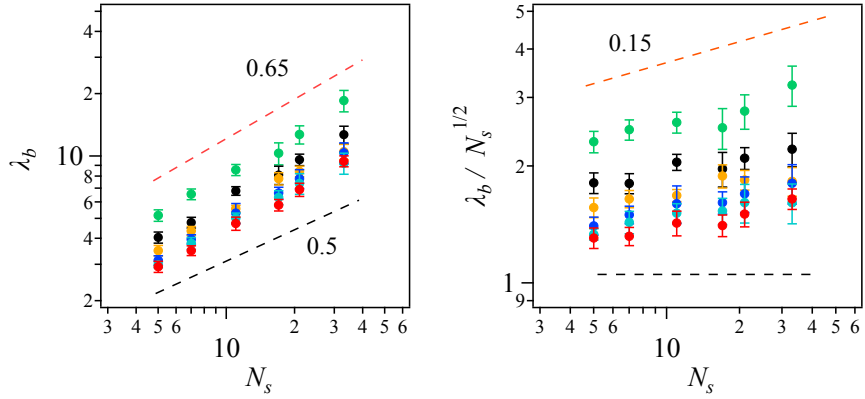

**Figure S5** Stretch at break $\lambda_b$ plotted against strand molecular weight $N_s = 2N_a + 1$ at $\varphi_c =0.95$ (a) and normalized by $N_s^{1/2}$ (b) for the simulations with non-linear spring constant $f_{ik} = (1 - \mathbf{b}_{ik}^2/b_{max}^2)^{-1}$ at $b_{max} = 2$. The $f$ values are 3 (green), 4 (black), 5 (orange), 6 (blue), 7(cyan), and 8 (red), respectively. Error bars indicate standard deviations for eight simulation runs. Black and red broken lines indicate power-law relations with the exponent of 0.5 and 0.65, respectively.

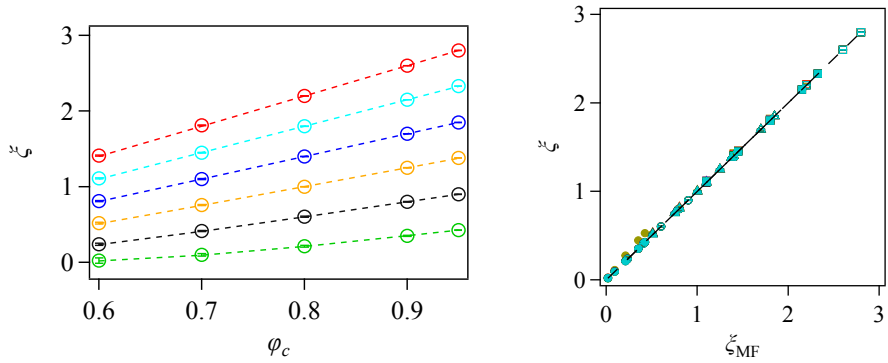

**Figure S6** Cycle rank per branch point $\xi$ plotted against $\varphi_c$ for various $f$ with $N_a$=5 (left), and $\xi$ plotted against that from the mean-field theory $\xi_{MF}$ (right). In the left panel, the $f$ values are 3 (green), 4 (black), 5 (orange), 6 (blue), 7(cyan), and 8 (red), respectively. Unfilled circles and broken curves indicate the simulation results and the theoretical prediction, respectively. In the right panel, $N_a$ values are 2 (yellow), 3 (red), 5 (blue), 8 (orange), 10 (green), 16 (black), and 20 (cyan). Symbols indicate $f = 3$ (filled circle), 4 (unfilled circle), 5 (filled triangle), 6 (unfilled triangle), 7 (filled square), and 8 (unfilled square). The broken line shows $\xi = \xi_{MF}$. Error bars indicate standard deviations for eight simulation runs, although most are smaller than the symbols.

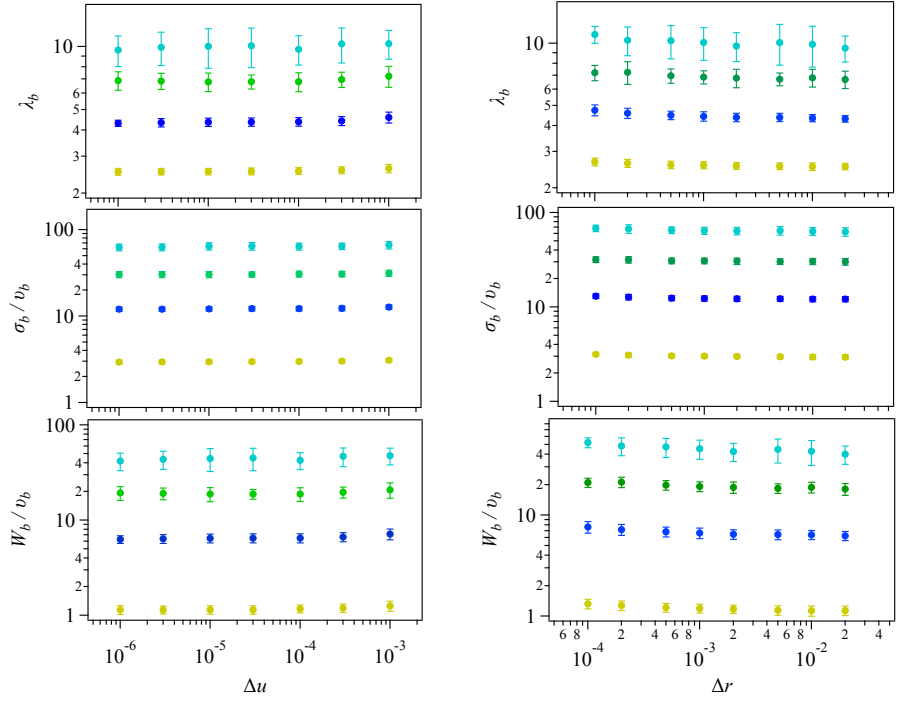

**Figure S7** Fracture characteristics $\lambda_b$, $\sigma_b/v_{br}$, and $W_b/v_{br}$ plotted against the simulation parameters for energy minimization $\Delta u$ (left) and $\Delta r$ (right) for $N_a = 2$ (yellow), 5 (blue), 10 (green), and 20 (cyan), respectively. $(f, \varphi_c) = (4, 0.95)$. Error bars indicate standard deviations for eight simulation runs.

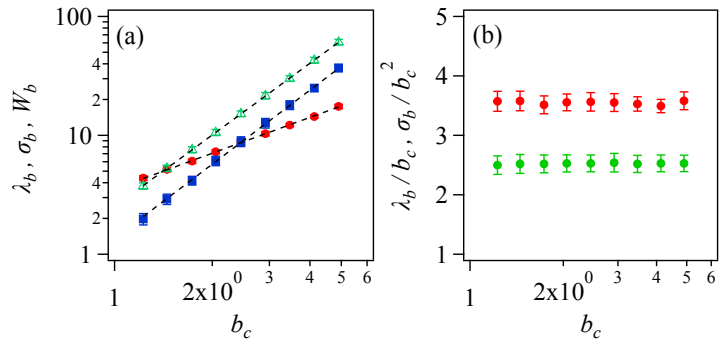

**Figure S8** Fracture characteristics $\lambda_b$, $\sigma_b$, and $W_b$ plotted against the simulation parameter for bond breakage $b_c$ (a). In Fig (b), $\lambda_b$ and $\sigma_b$ are normalized by $b_c$ and $b_c^2$. $(f, \varphi_c, N_a) = (4, 0.95, 5)$.

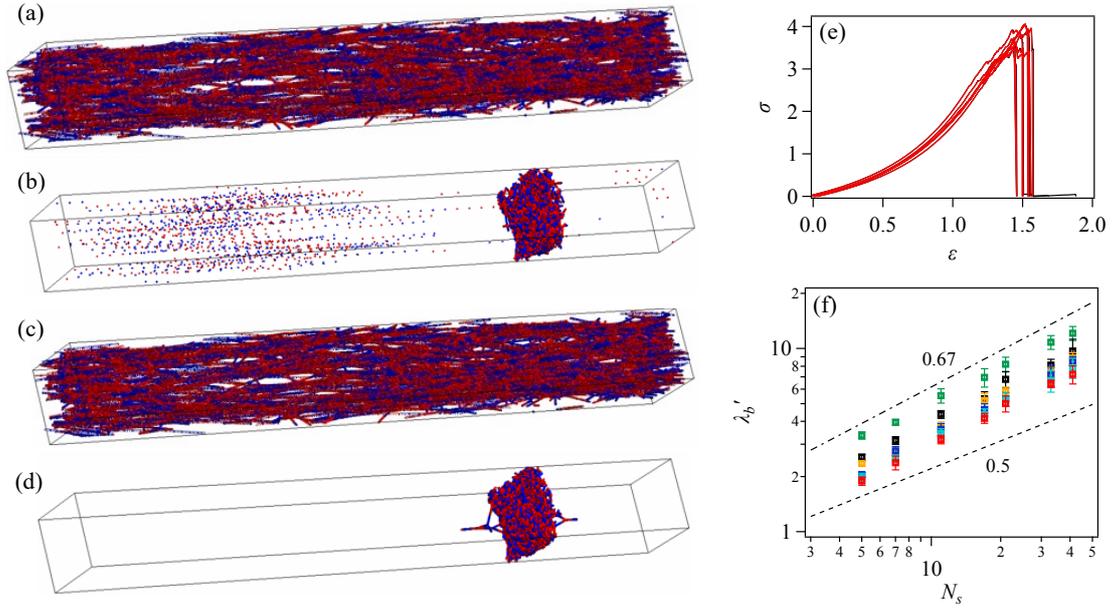

**Figure S9** Comparison between different simulation schemes with and without allowing network fragmentation. Panels (a) and (b) show snapshots for a network with $(N_a, f, \varphi_c) = (5, 4, 0.95)$ at $\varepsilon = 1.49$ (a), and that after the breakage at $\varepsilon = 1.50$ (b) for the simulation allowing network fragmentation. The dots are fragmented segments. Panels (c) and (d) show snapshots for the same network calculated by the scheme, disallowing the fragmentation. Panel (d) is the structure incompletely minimized to exhibit a few antlers on the relaxation. Panel (e) shows the stress-strain curves for eight independent simulations with $(N_a, f, \varphi_c) = (5, 4, 0.95)$, and the results with and without fragmentation are shown by black and red curves, respectively. Panel (f) shows the stretch at break from the simulation without fragmentation $\lambda_b'$ at $\varphi_c = 0.95$ for $f$ at 3 (green), 4 (black), 5 (orange), 6 (blue), 7(cyan), and 8 (red), respectively. Error bars indicate standard deviations for eight simulation runs. The dashed and broken lines indicate slopes of 0.67 and 0.5, respectively.

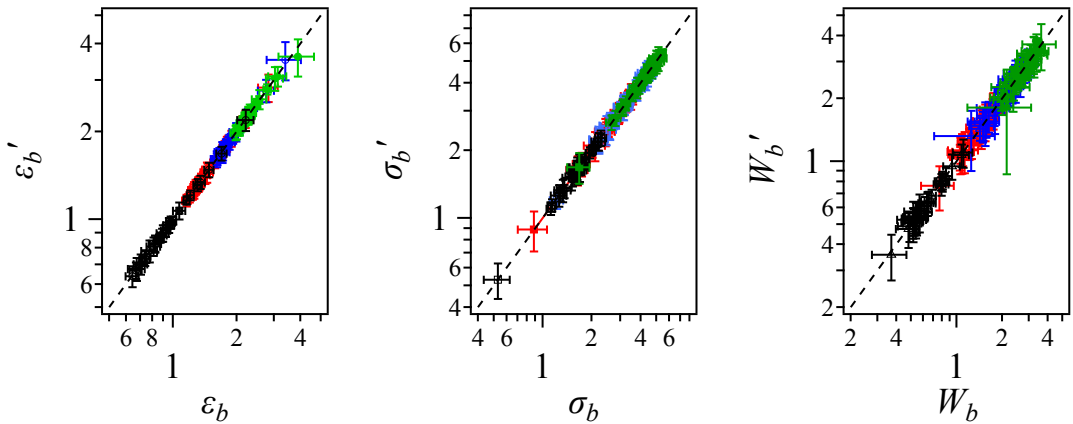

**Figure S10** Comparison of fracture characteristics between simulation schemes with and without network fragmentation. The stretch at break, stress at break, and work for fracture without fragmentation, $\lambda_b'$, $\sigma_b'$, and $W_b'$ are plotted against those without fragmentation, $\lambda_b$, $\sigma_b$, and $W_b$ for $N_a = 2$ (black), 5 (red), 10 (blue), and 20 (green), with $f = 3\sim 8$ and $\varphi_c = 0.6\sim 0.95$.

Error bars indicate standard deviations for eight simulation runs. Broken lines indicate the equivalency between compared values.

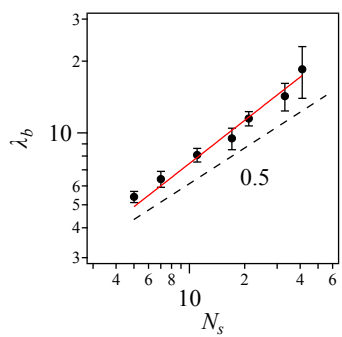

**Figure S11** Stretch at break obtained from Brownian simulations for the networks with $(f, \varphi_c) = (4, 0.95)$. The elongation rate was $10^{-4}$. The solid red line shows the best fit of the data toward a power law function with a slope of 0.61. The broken black line indicates a slope of 0.5. Error bars correspond to the standard deviation of eight independent simulation runs.